\documentclass[12pt]{iopart}

\begin{document}

% Hector Vargas Rodriguez
% Postal address:
% Manuel de Mimbela 1744
% Colonia Jardines del Country
% 44210 Guadalajara, Jal. Mexico
% Telephone: (33) 3824-2994

\input amssym.def

\input amssym

\title[Cylindrically symmetric non-aligned
Einstein--Maxwell solutions]{Cylindrically symmetric non-aligned
Einstein--Maxwell solutions with rotation and pseudorotation of
the types $O$ and $N$}

\author{H\'ector Vargas Rodr{\'\i}guez}
\address{Department of physics, C.U.C.E.I, Guadalajara University,
M\'exico. }

\ead{hv{\underline{ }}8@yahoo.com, hvargas@udgphys.intranets.com}

\begin{abstract}
A new solution to the Einstein--Maxwell field
equations is presented describing a cylindrically
symmetric homogeneous cosmology. The solution is
conformally flat, it possesses seven Killing
vectors of which the timelike one is rotating and
one of the spacelike, pseudorotating.  Our
solution also admits a Kerr-Schild form. It is
alternatively produced by different electromagnetic
sources some of which represent constant null
electromagnetic fields, while the others, a
circularly polarized plane electromagnetic wave
(seemingly, a unique situation in general
relativity). The concrete electromagnetic
four-potentials are found from the assumption that
they are proportional to the Killing covectors.
The general solution is obtained for timelike
and null geodesics. Finally, we find that this
space-time admits closed timelike non-geodesic lines.
\end{abstract}

\submitto{\CQG}

\pacs{0420.Jb, 0440.Nr, 0430.-w}

\section{Introduction}

A stationary space-time is that which admits
a temporal Killing vector $\xi_{[1]}=\partial_t$,
besides it is said to be
cylindrically symmetric if it is axisymmetric about an infinite
axis  and translationally invariant along that axis (here the $z$
axis); {\em i. e.} the space-time also admits the spatial Killing
vectors $\xi_{[2]}=\partial_z$, $\xi_{[3]}=\partial_\varphi$.
These three Killing vectors form an Abelian group $G_3$. We
suppose that both $\xi_{[1]}$ and $\xi_{[2]}$ are not
hypersurface--orthogonal,
\begin{equation}
*(\xi_{[1]}\wedge d\xi_{[1]})\neq 0,~~~~~~~~
*(\xi_{[2]}\wedge d\xi_{[2]})\neq 0;
\label{(1.1)}
\end{equation}
of course this means that the temporal Killing
vector $\xi_{[1]}$ is rotating, however since
$\xi_{[2]}$ is a spatial vector, we shall say
that it is `pseudorotating' and that the
corresponding space-time is `pseudostationary'
in this sense. Hence the line element becomes
(we suppose that in rotation and pseudorotation
one and the same function, $f$, is present)
\begin{equation}
ds^2=\hbox{\large e}^{2\alpha}(dt+f d \varphi)^2
-\hbox{\large e}^{2\beta}d \rho^2
-\hbox{\large e}^{2\delta}d\varphi^2
-\hbox{\large e}^{2\alpha}(dz+f d\varphi)^2,
\label{(1.2)}
\end{equation}
which depends only on $\rho$; although below we
first shall use Cartesian-like coordinates,
later the cylindrical-like coordinates are to be
revisited. One can find in the literature examples
of cylindrically symmetric Einstein-Maxwell fields
with rotation, like that of Wilson; or with
pseudorotation, like that of Chitre, see for
example Kramer {\em et al} [5]; but there are no
examples possessing the both properties at once.

In this paper we obtain an exact solution of the
self-consistent Einstein--Maxwell equations
having the both properties, rotation and
pseudorotation as well. Here a nonstandard
approach in solving the Einstein--Maxwell
equations is followed: Once the solution to
Einstein's equations is known, one proceeds
to find all its  Killing vectors. But then we
propose the ansatz that at least one of the
Killing (co)vectors is proportional to the
exact electromagnetic four-potential
(co)vector of this problem. It is revealed
that four of the seven Killing vectors in
fact satisfy the sourceless Einstein--Maxwell
equations. The ansatz is inspired by a conjecture
formulated by Horsk\'y and Mitskievich, see [3],[4].
This conjecture is a generalization of the fact
that the timelike Killing covectors of the
Schwarzschild and Kerr solutions coincide (up to
a constant factor) with the electromagnetic
four-potentials of the Reissner--Nordstr\"om and
Kerr--Newman fields respectively. The conjecture
states that {\em the electromagnetic four-potential
covector of a stationary self-consistent
Einstein--Maxwell field is simultaneously
proportional (up to a constant factor) to
the Killing covector of the corresponding vacuum
space-time when the parameter connected with the
electromagnetic field of the self consistent
problem is set equal to zero, this parameter
coinciding with the aforementioned constant factor}
(see Cataldo {\em et al} [1]).

In Section 2, we obtain the solution in both
Cartesian-like and cylindrical-like coordinates,
a natural tetrad basis being introduced for either
form of the metric, and the coordinate transformation
connecting the both forms are explicitly given.

In Section 3, the properties of our solution
are outlined, {\em i. e.} that it is of the
Petrov type $O$, moreover it possesses an
isometry group $G_7$, in addition it accepts
a Kerr-Schild form; besides, its electromagnetic
sources alternatively correspond to two different
kinds: either constant orthogonal electric and
magnetic fields, or a left circularly polarized
plane electromagnetic wave [in our space-time,
these two distintic alternatives lead to one and
the same stress-energy tensor (\ref{(2.12)});
this situation seems to be a unique in the
electromagnetic theory]. In all these cases
the electromagnetic fields are non aligned,
{\em i.e.} they posses other symmetries than that
of the metric field (which naturally coincides
with the symmetry of the electromagnetic
stress-energy tensor). Also the motion of
test particles and photons in this space-time
is studied. Finally,  it is shown that closed
timelike lines exist in this space-time.

In Section 4, not only the concluding remarks
are given, but an exact superposition of our
solution and the `anti-Mach' Ozv\'ath-Sch\"ucking
solution is also presented. These both results
(our solution and its superposition with the
`anti-Mach' one) are counter-examples to some
widely accepted views `prohibiting' such types
of exact solutions.

We denote derivative with respect to the
coordinate $x$ by a prime, and use the units
in which $c=1$. Four-dimensional indices
are Greek ones, the space-time signature
being ($+ - - \,- $), the Ricci tensor
$R_{\mu\nu}:= {R^{\lambda}}_{\mu\nu\lambda}$. The
symbol $*$ denotes the Hodge star as well as the
Levi-Civit\`a dual conjugation (when it is
written under or over a pair of corresponding
tensor indices).

\section{A stationary and pseudostationary solution}
Consider the line element (\ref{(1.2)}),
\[
ds^2=\hbox{\large e}^{2\alpha}(dt+f d y)^2
-\hbox{\large e}^{2\beta}d x^2
-\hbox{\large e}^{2\delta}d y^2
-\hbox{\large e}^{2\alpha}(dz+f d y)^2,
\]
where $\alpha$, $\beta$, $\delta$ and $f$ are functions of $x$
only. We introduce the natural orthonormalized basis related to
this metric, its covariant and contravariant forms being
respectively
\[
\theta^{(0)}=\hbox{\large e}^{\alpha}(dt+f\, d y),~~~
\theta^{(1)}=\hbox{\large e}^{\beta}d x,~~~
\theta^{(2)}=\hbox{\large e}^{\delta}d y,~~~
\theta^{(3)}=\hbox{\large e}^{\alpha}(dz+f\, d y),
\]
and
\[
X_{(0)}=\hbox{\large e}^{-\alpha}\partial_t,~
X_{(1)}=\hbox{\large e}^{-\beta}\partial_ x,~
X_{(2)}=\hbox{\large e}^{-\delta}(\partial_
y-f\partial_t-f\partial_z),~ X_{(3)}=\hbox{\large
e}^{-\alpha}\partial_z.
\]
Making use of Cartan's structure equations, it is easy to
calculate the Riemann tensor components and then those of the
Ricci curvature:
\[
R_{(0)(0)}=
-\left(\alpha'\hbox{\large e}^{2\alpha-\beta+\delta}
\right)'\hbox{\large e}^{-2\alpha-\beta-\delta}
-\frac{1}{2}{f'}^2\hbox{\large e}^{2(\alpha-\beta-\delta)},
\]
\[
R_{(1)(1)}=2\left(\alpha'\hbox{\large e}^{\alpha-\beta}
\right)'\hbox{\large e}^{-\alpha-\beta}+
\left(\delta'\hbox{\large e}^{\delta-\beta}
\right)'\hbox{\large e}^{-\delta-\beta},
\]
\[
R_{(2)(2)}=\left(\delta'\hbox{\large e}^{-\beta+\delta}
\right)'\hbox{\large e}^{-\beta-\delta}+
2\alpha'\delta'\hbox{\large e}^{-2\beta},
\]
\[
R_{(3)(3)}=\left(\alpha'\hbox{\large e}^{2\alpha-\beta+\delta}
\right)'\hbox{\large e}^{-2\alpha-\beta-\delta}-
\frac{1}{2}{f'}^2\hbox{\large e}^{2(\alpha-\beta-\delta)},
\]
\[
R_{(0)(2)}=-R_{(2)(3)}=-\frac{1}{2}\left(f'\hbox{\large
e}^{4\alpha-\beta-\delta} \right)'\hbox{\large
e}^{-3\alpha-\beta},
\]
\[
R_{(0)(3)}=
\frac{1}{2}{f'}^2\hbox{\large e}^{2(\alpha-\beta-\delta)},
\]
(all other Ricci components are equal to zero);
\[
\frac{1}{2}R=-
\left(\alpha'\hbox{\large e}^{2\alpha-\beta+\delta}
\right)'\hbox{\large e}^{-2\alpha-\beta-\delta}-
\left(\alpha'\hbox{\large e}^{\alpha-\beta}
\right)'\hbox{\large e}^{-\alpha-\beta}-
\left(\delta'\hbox{\large e}^{\alpha-\beta+\delta}
\right)'\hbox{\large e}^{-\alpha-\beta-\delta}.
\]
Then, the Einstein tensor components are
\[
G_{(0)(0)}=
\left(\alpha'\hbox{\large e}^{\alpha-\beta}
\right)'\hbox{\large e}^{-\alpha-\beta}+
\left(\delta'\hbox{\large e}^{\delta-\beta}
\right)'\hbox{\large e}^{-\delta-\beta}+
\alpha'\delta'\hbox{\large e}^{-2\beta}
-\frac{1}{2}{f'}^2\hbox{\large e}^{2(\alpha-\beta-\delta)},
\]
\[
G_{(1)(1)}=
-\alpha'\left(\alpha'+2'\delta'\right)\hbox{\large e}^{-2\beta},
\]
\[
G_{(2)(2)}=
-2\left(\alpha'\hbox{\large e}^{\alpha-\beta}
\right)'\hbox{\large e}^{-\alpha-\beta}-
{\alpha'}^2\hbox{\large e}^{-2\beta},
\]
\[
G_{(3)(3)}=
-\left(\alpha'\hbox{\large e}^{\alpha-\beta}
\right)'\hbox{\large e}^{-\alpha-\beta}-
\left(\delta'\hbox{\large e}^{\alpha-\beta+\delta}
\right)'\hbox{\large e}^{-\alpha-\beta-\delta}
-\frac{1}{2}{f'}^2\hbox{\large e}^{2(\alpha-\beta-\delta)},
\]
\[
G_{(0)(3)}=R_{(0)(3)},~~~~~~~~~~G_{(0)(2)}=-G_{(2)(3)}=R_{(0)(2)}.
\]
We shall consider the gravitational field with electromagnetic
sources, so that the scalar curvature vanishes, $R=0$. Assuming
that $G_{(1)(1)}=G_{(2)(2)}=G_{(0)(2)}=-G_{(2)(3)}=0$, one sees
immediately that $G_{(1)(1)}=G_{(2)(2)}=0$ are satisfied if
$\alpha=0$. Using the freedom on the choice of the $x$ coordinate
we put $\beta=0$. To satisfy $R=0$, we have two choices, the first
one, $\delta=0$, takes us to Cartesian-like coordinates, and the
second one, $\delta=\ln (x)$, to cylindrical-like coordinates,
hence reinterpreting the coordinates $t$, $x$, $y$, $z$ as $\tilde
t$, $\rho$, $\varphi$, $\tilde z$. The both result in the same
solution. Finally, $G_{(0)(2)}=-G_{(2)(3)}=0$ is fulfilled in the
first case if $f(x)=Cx$, and in the second case if
$f(\rho)=\frac{C}{2}\rho^2$. Thus the solution in Cartesian-like
coordinates is
\begin{equation}
ds^2=(dt+C x d y)^2-d x^2-d y^2-(dz+C x d y)^2,
\label{(2.7)}
\end{equation}
with the natural tetrad
\[
\theta^{(0)}=dt+Cx\, d y,~~~~
\theta^{(1)}=d x,~~~~
\theta^{(2)}=d y,~~~~
\theta^{(3)}=dz+Cx \, d y.
\]
In cylindrical-like coordinates,
\begin{equation}
ds^2=\left(d\tilde t+\frac{C}{2}\rho^2 d\varphi\right)^2-
d\rho^2-\rho^2 d\varphi^2-\left(d\tilde z+\frac{C}{2}\rho^2
d\varphi\right)^2,
\label{(2.9)}
\end{equation}
with
\[
\tilde\theta^{(0)}=d\tilde t+\frac{C}{2}\rho^2 \, d\varphi,~~~~
\tilde\theta^{(1)}=d \rho,~~~~
\tilde\theta^{(2)}=\rho d \varphi,~~~~
\tilde\theta^{(3)}=d\tilde z+\frac{C}{2}\rho^2 \, d\varphi.
\]
The both forms are related via the transformation
\[
t=\tilde t-\frac{C}{2}xy,~~~~z=\tilde z-\frac{C}{2}xy,~~~~
x=\rho\cos\varphi,~~~~y=\rho\sin\varphi.
\]
In the both cases the electromagnetic energy-momentum tensor
(EEMT) is
\[
T=-\frac{1}{\varkappa}G=\frac{C^2}{2\varkappa}
(\theta^{(0)}-\theta^{(3)}) \otimes(\theta^{(0)}-\theta^{(3)})=
\frac{C^2}{2\varkappa}(dt-dz) \otimes(dt-dz)=
\]
\begin{equation}
=\frac{C^2}{2\varkappa}(\tilde \theta^{(0)} -\tilde \theta^{(3)})
\otimes(\tilde\theta^{(0)}- \tilde\theta^{(3)})=
\frac{C^2}{2\varkappa}(d\tilde t- d\tilde z) \otimes(d\tilde
t-d\tilde z), \label{(2.12)}
\end{equation}
$\varkappa$ being the gravitational constant.
The nature of the electromagnetic sources is
discussed in the subsection 3.2. Here it is worth
being mentioned that (\ref{(2.12)}) has the canonical
structure for a null electromagnetic field
({\em cf.} Synge [10]).

\section{Properties of the new solution}

\subsection{Geometry}

\subsubsection{Petrov type.}
A calculation immediately shows that in the
space-time (\ref{(2.7)}), (\ref{(2.9)}) the Weyl
tensor vanishes, thus it is of the Petrov type $O$.
Of course this means that the metric is conformally flat
and that the (Ricci-flat) Riemann tensor could
be expressed exclusively in terms of the sources
\begin{equation}
R_{(\alpha)(\beta)(\gamma)(\delta)}=
\varkappa\left(
T_{(\alpha)[(\gamma)}\eta_{(\delta)](\beta)}-
T_{(\beta)[(\gamma)}\eta_{(\delta)](\alpha)}
\right).
\label{(3.1.1)}
\end{equation}

\subsubsection{Killing vectors.}
Our solution admits isometry group $G_7$, the
corresponding Killing vectors in  contravariant
form using Cartesian-like coordinates are
\[
\xi_{[1]}=\partial_t,
\]
\[
\xi_{[2]}=\partial_z,
\]
\[
\xi_{[3]}=\frac{C}{2}( y^2-x^2)(\partial_t +\partial_z)-y
\partial_x + x \partial_ y,
\]
\[
\xi_{[4]}=\partial_ y,
\]
\[
\xi_{[5]}=C y(\partial_t+\partial_z)-\partial_ x,
\]
\[
\xi_{[6]}=C x\sin[C(t-z)](\partial_t+\partial_z)-
\cos[C(t-z)]\partial_ x -\sin[C(t-z)]\partial_ y,
\]
\[
\xi_{[7]}=-C x\cos[C(t-z)](\partial_t+\partial_z)-
\sin[C(t-z)]\partial_ x
+\cos[C(t-z)]\partial_ y,
\]
and the covariant ones,
\[
\xi_{[1]}=dt+C x d y,
\]
\[
\xi_{[2]}=-dz-C x d y,
\]
\[
\xi_{[3]}=\frac{C}{2}(x^2+y^2)(dt-dz)+y d x - x d y,
\]
\[
\xi_{[4]}=C x(dt-dz)-d y,
\]
\[
\xi_{[5]}=C y(dt-dz)+d x,
\]
\[
\xi_{[6]}=\cos[C(t-z)]d x+\sin[C(t-z)]d y,
\]
\[
\xi_{[7]}=\sin[C(t-z)]d x-\cos[C(t-z)]d y.
\]
The corresponding contravariant forms in
cylindrical-like coordinates are
\[
\xi_{[1]}=\partial_{\tilde t},
\]
\[
\xi_{[2]}=\partial_{\tilde z},
\]
\[
\xi_{[3]}= \partial_\varphi,
\]
\[
\xi_{[4]}=\frac{C}{2}\rho\cos(\varphi)(\partial_{\tilde
t}+\partial_{\tilde z})+\sin(\varphi)\partial_\rho-
\frac{1}{\rho}\cos(\varphi)\partial_\varphi,
\]
\[
\xi_{[5]}=\frac{C}{2}\rho\sin(\varphi)(\partial_{\tilde
t}+\partial_{\tilde z})-\cos(\varphi)\partial_\rho+
\frac{1}{\rho}\sin(\varphi)\partial_\varphi,
\]
\[
\xi_{[6]}=
-\sin[C(\tilde t-\tilde z)-\varphi]
\left(\frac{1}{2}C\rho(\partial_{\tilde t}+\partial_{\tilde z})
+\frac{1}{\rho}\partial_\varphi\right)-
\cos[C(\tilde t-\tilde z)-\varphi]\partial_\rho,
\]
\[
\xi_{[7]}=-\cos[C(\tilde t-\tilde z)-\varphi]
\left(\frac{1}{2}C\rho(\partial_{\tilde t}+\partial_{\tilde z})
-\frac{1}{\rho}\partial_\varphi\right)-
\sin[C(\tilde t-\tilde z)-\varphi]\partial_\rho,
\]
and the covariant forms,
\[
\xi_{[1]}=d\tilde t+\frac{C}{2}\rho^2 d\varphi,
\]
\[
\xi_{[2]}=-d\tilde z-\frac{C}{2}\rho^2 d\varphi,
\]
\[
\xi_{[3]}=\frac{C}{2}\rho^2(d\tilde t-d\tilde z)-
\rho^2 d\varphi,
\]
\[
\xi_{[4]}=C\rho\cos(\varphi)(d\tilde t-d\tilde
z)-\sin(\varphi)d\rho- \rho\cos(\varphi)d\varphi,
\]
\[
\xi_{[5]}=C\rho\sin(\varphi)(d\tilde t-d\tilde
z)+\cos(\varphi)d\rho- \rho\sin(\varphi)d\varphi,
\]
\[
\xi_{[6]}=\cos[C(\tilde t-\tilde z)-\varphi]d\rho+
\rho\sin[C(\tilde t-\tilde z)-\varphi]d\varphi,
\]
\[
\xi_{[7]}=\sin[C(\tilde t-\tilde z)-\varphi]d\rho-
\rho\cos[C(\tilde t-\tilde z)-\varphi]d\varphi.
\]
Contravariant Killing vectors have the following non
trivial commutators:
\[
\begin{array}{l}
{[\xi_{[1]},\xi_{[6]}]=-C\xi_{[7]}},    \\
{[\xi_{[1]},\xi_{[7]}]=C\xi_{[6]}}, \\
{[\xi_{[2]},\xi_{[6]}]=C\xi_{[7]}}, \\
{[\xi_{[2]},\xi_{[7]}]=-C\xi_{[6]}},   \\
{[\xi_{[3]},\xi_{[4]}]=-C\xi_{[5]}},
\end{array}
~~~~~~~~
\begin{array}{l}
{[\xi_{[3]},\xi_{[5]}]=-C\xi_{[4]}},            \\
{[\xi_{[3]},\xi_{[6]}]=C\xi_{[7]}},         \\
{[\xi_{[3]},\xi_{[7]}]=-C\xi_{[6]}},            \\
{[\xi_{[4]},\xi_{[5]}]=C(\xi_{[1]}+\xi_{[2]})}, \\
{[\xi_{[6]},\xi_{[7]}]=C(\xi_{[1]}+\xi_{[2]})}.
\end{array}
\]
Only $\xi_{[1]}$ is timelike and all others are space-like,
\[
\xi_{[1]}\cdot\xi_{[1]}=1,~~~~~~
\xi_{[3]}\cdot\xi_{[3]}=-\rho^2,
\]
\[
\xi_{[2]}\cdot\xi_{[2]}=\xi_{[4]}\cdot\xi_{[4]}=
\xi_{[5]}\cdot\xi_{[5]}=\xi_{[6]}\cdot\xi_{[6]}
=\xi_{[7]}\cdot\xi_{[7]}=-1,
\]
Finally we could see that $\xi_{[1]}$ is rotating and
$\xi_{[2]}$ is pseudorotating,
\[
\omega=\frac{1}{2}*(\xi_{[1]}\wedge d
\xi_{[1]})=\frac{C}{2}\tilde\theta^{(3)}, ~~~~~~
\varpi=\frac{1}{2}*(\xi_{[2]}\wedge d
\xi_{[2]})=\frac{C}{2}\tilde\theta^{(0)},
\label{(3.1.33)}
\]
here $\omega$ being the usual space-time angular
velocity which is directed along the negative $z$
axis, $\varpi$ is the space-time pseudo-angular-velocity
(introduced as an analogue of $\omega$), it is directed
along the $t$ axis; for the both, it could be seen
that if $C$ changes its sign, they change their
directions.

It is interesting that $ds^2$ can be written exclusively
in terms of the Killing vectors, for example (\ref{(2.9)})
is equivalent to
\begin{equation}
ds^2=\xi_{[1]}\xi_{[1]}-\xi_{[2]}\xi_{[2]}
-\xi_{[6]}\xi_{[6]}-\xi_{[7]}\xi_{[7]}
\label{(3.1.34)}
\end{equation}

\subsubsection{Kerr-Schild form of the metric}
If the metric (\ref{(2.9)}) is written as
\[
ds^2=d\tilde t^2- d\rho^2-d\tilde z ^2 +
\rho^2d[C(\tilde t-\tilde z)-\varphi] d\varphi
\]
and one performs the change
\[
\varphi=\tilde\varphi+\frac{C}{2}(\tilde t-\tilde z),
\]
one arrives to the Kerr--Schild form
\begin{equation}
ds^2=d\tilde t^2- d\rho^2-\rho^2 d\tilde\varphi ^2 -d\tilde z ^2+
\frac{C^2}{4}\rho^2(d\tilde t-d\tilde z)^2.
\label{(3.1.37)}
\end{equation}
It is easily checked that $d\tilde t-d\tilde z$
is a null (co)vector from the viewpoint of the
both Minkowski and the stationary and pseudostationary
metric (\ref{(3.1.37)}), and that this vector
is geodesic. Contrary to the Kerr congruence
this congruence does not rotate; this may suggest
that not all stationary space-times correspond to
null rotating congruences or that the pseudorotation
somehow compensates the rotation of the null congruence.

\subsection{The electromagnetic field}
Modifying the conjecture proposed by Horsk\'y and
Mitskievich, we shall now look for electromagnetic
four-potentials proportional to the Killing covectors
(below their Cartesian forms being used),
\begin{equation}
A=k\xi
\label{(3.2.1)}
\end{equation}
(the electromagnetic field tensor is its exterior
derivative, $F=dA$) and check if they would satisfy
the vacuum Maxwell equations, in our space-time,
\[
(\sqrt{-g}F^{\mu\nu})_{,\nu}=0,
\]
and lead to one and the same form of the EEMT
(\ref{(2.12)}). This will put a constraint on the
proportionality constant $k$. It is found that four
of these Killings $\xi_{[4]}$, $\xi_{[5]}$, $\xi_{[6]}$,
$\xi_{[7]}$, and the linear superposition $\xi_{[4]}\pm
\xi_{[5]}$, fulfill the above requirements. So we have
five cases. [In fact, $\xi_{[3]}$ satisfies the vacuum
Maxwell equations, but it yields the EEMT different
from (\ref{(2.12)}), thus this candidate of the
four-potential  describes a merely test electromagnetic
field in our space-time. However it could serve as the
first step in a further generalization of our solution.]

In order to describe the electric and magnetic field
vectors we shall introduce a reference frame described
by the monad (see Mitskievich [9]),
\begin{equation}
\tau= \theta^{(0)}=d t+C x dy.
\label{(3.2.3)}
\end{equation}This reference frame is rotating
\begin{equation}
\omega=\frac{1}{2}*(\tau\wedge d\tau)=\frac{C}{2} \theta^{(3)}.
\label{(3.2.4)}
\end{equation}
but has neither acceleration
\begin{equation}
G=-*(\tau\wedge*d\tau)=0,
\label{(3.2.5)}
\end{equation}
nor expansion and shear since the rate-of-strain tensor
vanishes
\begin{equation}
D_{\mu\nu}=\frac{1}{2}\pounds_\tau b_{\mu\nu}=0,
\label{(3.2.6)}
\end{equation}
here $b_{\mu\nu}=g_{\mu\nu}-\tau_\mu\tau_\nu$ is the three metric
in the local subspace orthogonal to the monad and $\pounds_\tau$
denotes the Lie derivative with respect to the monad which
coincides with $\xi_{[1]}$.

With respect to this reference frame we split
the electromagnetic field tensor in the electric
and magnetic (co)vectors
\begin{equation}
E=*(\tau\wedge *F),~~~~~~~~~~B=*(\tau\wedge F).
\label{(3.2.7)}
\end{equation}
the electromagnetic momentum density (Poynting
covector, $c=1$) being ({\em cf.} [9])
\begin{equation}
S=\frac{1}{4\pi}*(E \wedge \tau\wedge B).
\label{(3.2.8)}
\end{equation}
In all the five cases below it is found that
the Poynting covector is
\begin{equation}
S= -\frac{C^2}{2\varkappa}\theta^{(3)},
\label{(3.2.9)}
\end{equation}
as it could be seen from the EEMT; it is directed
along the positive $z$ axis and does not depend on
the sign of $C$. Also for the all five cases below
\[
F_{(\mu)(\nu)}F^{(\mu)(\nu)}=0,~~~~
F_{(\mu)(\nu)}F\stackrel{(\mu)(\nu)}{*}=0.
\]
Since these both electromagnetic field invariants
vanish, we come to the pure null type electromagnetic
field.

The first three cases, $\xi_{[4]}$, $\xi_{[5]}$ and
$\xi_{[4]}\pm\xi_{[5]}$ correspond to a constant
orthogonal electric and magnetic fields, while the
last two cases, $\xi_{[6]}$ and $\xi_{[7]}$, correspond
to a left circularly polarized (positive helicity)
plane electromagnetic {\em wave} propagating in the
direction of the positive $z$ axis. Hence the
plane electromagnetic wave has its spin angular momentum
in an opposite direction to that of the angular velocity
of the space-time. If $C$ changes its sign, then the plane
electromagnetic wave acquires negative helicity and the
relative situation continues to be as before.

In the following subsubsections, the reader  will see that
the alternative variants of electromagnetic field are always
non aligned with the space-time  geometry, though the (one
and the same)  stress-energy tensor does exactly correspond
to the space-time symmetry.

\subsubsection{The fourth Killing vector case}
The electromagnetic four-potential and field tensor are
\[
A_{[4]}=\sqrt{\frac{2\pi}{\varkappa}}C x(dt-dz),~~~~~~
F_{[4]}=\sqrt{\frac{2\pi}{\varkappa}}
C\theta^{(1)}\wedge\left(\theta^{(0)}- \theta^{(3)}\right).
\]
with the corresponding electric and magnetic covectors
\[
E_{[4]}=\sqrt{\frac{2\pi}{\varkappa}}C\theta^{(1)},
~~~~
B_{[4]}=\sqrt{\frac{2\pi}{\varkappa}}C\theta^{(2)}.
\]
\subsubsection{The fifth Killing vector case}
\[
A_{[5]}=\sqrt{\frac{2\pi}{\varkappa}}C y(dt-dz),~~~~~~
F_{[5]}=\sqrt{\frac{2\pi}{\varkappa}}
C\theta^{(2)}\wedge\left(\theta^{(0)}- \theta^{(3)}\right).
\]
\[
E_{[5]}=\sqrt{\frac{2\pi}{\varkappa}}C\theta^{(2)},
~~~~
B_{[5]}=-\sqrt{\frac{2\pi}{\varkappa}}C\theta^{(1)}.
\]
\subsubsection{A linear superposition of the fourth and
fifth Killing vector cases}
\[
A_{[4,5]}=\sqrt{\frac{\pi}{\varkappa}}C(x\pm y)(dt-dz),
\]
\[
F_{[4,5]}=\sqrt{\frac{\pi}{\varkappa}}C (\theta^{(1)} \pm
\theta^{(2)})\wedge\left(\theta^{(0)}-\theta^{(3)}\right).
\]
\[
E_{[4,5]}=\sqrt{\frac{\pi}{\varkappa}}C(\theta^{(1)}\pm \theta^{(2)}),
~~~~
B_{[4,5]}=\sqrt{\frac{\pi}{\varkappa}}C(\theta^{(2)}\mp \theta^{(1)}).
\]
\subsubsection{The sixth Killing vector case}
\[
A_{[6]}=\sqrt{\frac{2\pi}{\varkappa}}\left
\{\cos[C(t-z)]dx+\sin[C(t-z)]dy\right\},
\]
\[
F_{[6]}=\sqrt{\frac{2\pi}{\varkappa}}C\left\{
\sin[C(t-z)]\theta^{(1)}-\cos[C(t-z)]\theta^{(2)}\right\}
\wedge\left(\theta^{(0)}-\theta^{(3)}\right),
\]
\[
E_{[6]}=\sqrt{\frac{2\pi}{\varkappa}}C\left\{\sin[C(t-z)]\theta^{(1)}-
\cos[C(t-z)]\theta^{(2)} \right\},
\]
\[
B_{[6]}=-\sqrt{\frac{2\pi}{\varkappa}}C\left\{
\cos[C(t-z)]\theta^{(1)}-
\sin[C(t-z)]\theta^{(2)} \right\}.
\]
\subsubsection{The seventh Killing vector case}
This case could be obtained from the later, since if one changes,
for example, $z$ by $z+\pi/(2C)$ in $A_{[6]}$, it becomes
$A_{[7]}$. We come to the electromagnetic four-potential
\[
A_{[7]}=\sqrt{\frac{2\pi}{\varkappa}}
\left\{\sin[C(t-z)]dx-\cos[C(t-z)]dy\right\}
\]
and the field tensor
\[
F_{[7]}=\sqrt{\frac{2\pi}{\varkappa}}C
\left(\theta^{(0)}-\theta^{(3)}\right)\wedge\left\{
\cos[C(t-z)]\theta^{(1)}+
\sin[C(t-z)]\theta^{(2)}
\right\},
\]
while the corresponding electric and magnetic covectors are
\[
E_{[7]}=-\sqrt{\frac{2\pi}{\varkappa}}C
\left\{\cos[C(t-z)]\theta^{(1)}+
\sin[C(t-z)]\theta^{(2)} \right\},
\]
\[
B_{[7]}=\sqrt{\frac{2\pi}{\varkappa}}C\left\{
\sin[C(t-z)]\theta^{(1)}-
\cos[C(t-z)]\theta^{(2)} \right\}.
\]
When $C>0$, their contravariant forms are interpreted as
pertaining to a left circularly polarized electromagnetic wave (a
wave with positive helicity).

\subsection{Timelike and null geodesics}
From the geodesic line equation
\[
\frac{d}{d\lambda}\left(g_{\mu\nu}
\frac{dx^\nu}{d\lambda}\right)=\frac{1}{2}
g_{\alpha\beta,\mu}\frac{dx^\alpha}{d\lambda}
\frac{dx^\beta}{d\lambda},
\]
and the metric (\ref{(2.7)}), the first integrals follow
\[
\mu=0:~~~~\frac{dt}{d\lambda}+C x\frac{d y}{d\lambda}=K_0,
\]
\[
\mu=2:~~~~C x\left(\frac{dt}{d\lambda}-
\frac{dz}{d\lambda}\right)-\frac{d y}{d\lambda}=K_2,
\]
\[
\mu=3:~~~~\frac{dz}{d\lambda}+C x\frac{d y}{d\lambda}=K_3.
\]
alongside with the equation
\[
\mu=1:~~~~\frac{d^2 x}{d\lambda^2}+
C\left(\frac{dt}{d\lambda}-
\frac{dz}{d\lambda}\right)\frac{d y}{d\lambda}=0.
\]
We obtain the solution to the above equations, and then insert it
in $ds^2$,
\[
\left(\frac{ds}{d\lambda}\right)^2=
\left(\frac{dt}{d\lambda}+C x\frac{d y}{d\lambda}\right)^2-
\left(\frac{d x}{d\lambda}\right)^2-
\left(\frac{dz}{d\lambda}+C x\frac{d y}{d\lambda}\right)^2-
\left(\frac{d y}{d\lambda}\right)^2=\eta ,
\]
since it is already a first integral of the equations of motion.
Here we take $\eta=1$ for timelike geodesics (which describe the
motion of test massive particles) or $\eta=0$ for null geodesics
(which describe the motion of test photons), so that timelike and
null geodesics have the following parametric expressions:
\[
t(\lambda)=\left[\frac{\eta+(K_0-K_3)^2}
{2(K_0-K_3)}\right]\lambda
-\frac{K_0 ^2-K_3 ^2-\eta}{4C(K_0-K_3)^2}
\sin[2C(K_0-K_3)\lambda],
\]
\[
x(\lambda)=\sqrt{\frac{K_0 ^2-K_3 ^2-\eta}
{C^2(K_0-K_3)^2}}\cos[C(K_0-K_3)\lambda],
\]
\[
y(\lambda)=\sqrt{\frac{K_0 ^2-K_3 ^2-\eta}
{C^2(K_0-K_3)^2}}\sin[C(K_0-K_3)\lambda],
\]
\[
z(\lambda)=\left[\frac{\eta-(K_0-K_3)^2}
{2(K_0-K_3)}\right]\lambda
-\frac{K_0 ^2-K_3 ^2-\eta}
{4C(K_0-K_3)^2}\sin[2C(K_0-K_3)\lambda];
\]
here $K_0$ is always positive since it is usually
interpreted as energy per unit rest mass, in fact
$K_0\geq\sqrt{\eta+K_3 ^2}$. For test massive
particles the canonical parameter $\lambda$ is
interpreted as the proper time along the particle's
world line. Some of the integration constants were
put equal to zero by performing a translation in
$\lambda$, or by choosing the origin of the coordinates.

For test massive particles, $\eta=1$, we could see
that they could be at rest on the $z$-axis, or moving
along it, but if a test massive particle is outside
the $z$-axis, it revolves around the $z$-axis against
the space-time rotation while this particle is travelling
along the $z$-axis (when $C>0$).

For test photons, $\eta=0$, we could see that
they can travel only in the negative $z$-axis
direction, and they revolve against the
space-time rotation (when $C>0$).

\subsection{The closed-circuit journey along a timelike line}
In another form, the metric (\ref{(2.9)}) reads
\[
ds^2=(d\tilde t-d\tilde z)\left(d\tilde t+d\tilde z+C
\rho^2d\varphi\right)-d\rho^2-\rho^2d\phi^2.
\]
We divide the closed-circuit journey into two parts, the both
corresponding to $d\rho=0$:
\[
\mbox{(A)} ~~~ \left.\frac{d\tilde t}{ds}\right|_{{\mathrm A}}
=-\alpha<0, ~~~ \left.\frac{d\tilde z}{ds}\right|_{{\mathrm A}}
=\beta>0, ~~~ \left.\frac{d\varphi}{ds}\right|_{{\mathrm A}}
=\gamma\neq 0,
\]
and
\[
\mbox{(B)} ~~~ \left.\frac{d\tilde t}{ds}\right|_{{\mathrm B}}
=a>0, ~~~ \left.\frac{d\tilde z}{ds}\right|_{{\mathrm B}}
=-b<0, ~~~ \left.\frac{d\varphi}{ds}\right|_{{\mathrm B}}
=0.
\]
At any stage, the constants $\alpha, ~ \beta, ~
\gamma, ~ a$ and $b$ are subject to the constraint
\[
\left(\frac{d\tilde t}{ds}-\frac{d\tilde z}{ds}\right)\left(
\frac{d\tilde t}{ds}+\frac{d\tilde z}{ds}+C\rho^2
\frac{d\varphi}{ds}
\right)-\rho^2\left(\frac{d\varphi}
{ds}\right)^2=1.
\]
All this guarantees that the parameter $s$ grows
monotonically along the whole circuit, while the time
coordinate changes its behaviour from diminishing in A
to growth in B. In fact, in A we have
\begin{equation}
(\alpha+\beta) (\alpha-\beta-C\rho^2\gamma)
-\rho^2\gamma^2=1
\label{(3.4.5)}
\end{equation}
and in B, simply $a^2-b^2=1.$ The last relation is
equivalent to $a=\cosh\psi,~ b=\sinh\psi$.
Let us also suppose that
\[
\int_{{\mathrm A}}d\tilde t+
\int_{{\mathrm B}}d\tilde t=0,~~~~~~
\int_{{\mathrm A}}d\tilde z+
\int_{{\mathrm B}}d\tilde z=0,~~~~~~
\int_{{\mathrm A}}d\varphi=-2\pi
\]
respectively,
\begin{equation}
-\alpha s_{{\mathrm A}}+as_{{\mathrm B}}=0,~~~~~~
\beta s_{{\mathrm A}}-bs_{{\mathrm B}}=0,  ~~~~~~
\gamma s_{{\mathrm A}}= -2\pi,
\label{(3.4.9)}
\end{equation}
$s_{{\mathrm A}}$ and $s_{{\mathrm B}}$ being the
proper-time durations of the two parts of journey,
the minus sign in the $\varphi$
integral guarantees that $s_{{\mathrm A}}>0$.

When $s_{{\mathrm A}}=s_{{\mathrm B}}$,
then $\alpha=a$ and
$\beta=b$, and (\ref{(3.4.5)}) yields
\[
\gamma=-C(\alpha+\beta)=-C\hbox{\large e}^\psi.
\]
It is interesting to observe that the sign of $\gamma$ is
automatically the opposite to that of $C$ (the azimuthal
motion follows the space-time rotation, and this motion
is clearly non-geodesic).

From the last integral in (\ref{(3.4.9)}), we
found the proper-time durations
\[
s_{{\mathrm A}}=s_{{\mathrm B}}=
\frac{2\pi}{C}\hbox{\large e}^{-\psi}.
\]

\section{Concluding remarks}

One could find in Kramer {\em et al} [5], in table
33.5, a conclusion that conformally flat solutions
with non-aligned null electromagnetic fields do not
exist. The solution presented above is however a
counterexample. This conclusion in [5] is based on
the theorem 7.4 (in the same book), if the
electromagnetic null field is supposed to be only
a test one (as this usually occurs in the case when
the four-potential is simultaneously, up to a
constant factor, a Killing vector of the space-time
under consideration). But we have shown that here
we have a self-consistent solution of the
Einstein-Maxwell equations. See also in [5] the
theorem 28.7 which should correspond to principally
non existing solutions according to the same table 33.5.

The abundance of isometries in our solution suggests to consider
it as describing an electromagnetic homogeneous cosmological
model. This model is stationary (and pseudostationary), and its
curvature (the Riemann tensor) completely reduces to the
electromagnetic stress-energy tensor [see (\ref{(3.1.1)})]: one
may say that there is no free (intrinsic) gravitational field in
this space-time. The overall simplicity of this solution permits
to easily perform a complete integration of the geodesic equation
in its space-time in the general case. What is even stranger,
there are several equally good candidates for the concrete
electromagnetic fields as sources of its space-time geometry (the
reader would willingly accept the `candidates' mutually related by
a dual conjugation or dual rotation which do not change the
electromagnetic stress-energy tensor at all, but here we have so
different fields as, for example, a plane monochromatic
electromagnetic wave with the frequency $C$, and a pair of
time-independent electric and magnetic fields, a zero-frequency
wave, if one would wish to call it so). It is worth mentioning
that the four-potentials of these fields are also the Killing
vectors of this space-time. We have also seen that four of the
seven Killing vectors can be used in a very simple representation
of our solution (\ref{(3.1.34)}).

Like some other space-times, our solution permits
existence of closed timelike (non-geodesic) paths
({\em cf.} this property of the G\"odel cosmological
solution [2]).

It is easy to see that the Ozv\'ath--Sch\"ucking `anti-Mach'
solution (10.13), written in non-null coordinates, in [5] permits
an exact superposition with our solution. In the Kerr--Schild
form, this superposition is simply
\[
ds^2=dt^2-d\rho^2-\rho^2 d\varphi^2-dz^2-
(H+\frac{C^2}{4}\rho^2)(dt-dz)^2,
\]
with
\[
H(t,\rho,\varphi,z) = A\rho^2\cos(2B(t-z)+2\varphi).
\]
Our expressions for the electromagnetic field and the Ricci tensor
do not change their forms in the new orthonormal basis, the Petrov
type is now $N$. The possibility of such superposition is based on
the fact that the both fields, that of the Oszv\'ath--Sch\"ucking
wave and our electromagnetic field with its Poynting vector,
propagate in one and the same direction with the fundamental
velocity $c=1$, thus without any interaction on the Minkowski
background of the Kerr--Schild metric ({\em cf} [6,7]); the same
property of no interaction takes place for any objects (both
particles and fields, not only electromagnetic ones) with the same
type of propagation.

\ack

I would like to thank Prof. N. V. Mitskievich (thesis director)
for comments and helpful remarks on this work, which is
part of my Ph. D. Thesis research. I also thank CONACyT-M\'exico
for the scholarship grant No. 91290.

\References

\item[{[1] }] Cataldo M, Kumaradtya K K
and Mitskievich N V 1994
{\it Gen. Relat. and Grav.} {\bf 26} 847

\item[{[2] }] G\"odel K 1949
\RMP {\bf 21} 447

\item[{[3] }] Horsky J and Mitskievich N V 1989
{\it Czech. J. Phys.} {\bf B 39} 957.

\item[{[4] }] Horsky J and Mitskievich N V 1990
\CQG {\bf 7} 1523

\item[{[5] }] Kramer D, Stephani H, MacCallum M,
and Herlt E 1980 {\it Exact Solutions of
Einstein's Field Equations}
(Cambridge, UK: Cambridge University Press)

\item[{[6] }] Kumaradtya K K and
Mitskievich N V 1989
{\it J. Math. Phys.} {\bf 30} 1095

\item[{[7] }] Mitskievich N V 1981
{\it Experim. Technik der Physik} {\bf 29} 213

\item[{[8] }] Mitskievich N V and Tsalakou G A 1991
\CQG {\bf 8}, 209.

\item[{[9] }] Mitskievich N V 1996
Relativistic Physics in
Arbitrary Reference Frames {\it Preprint} gr-qc/9606051

\item[{[10] }] Synge J L 1965
{\it Relativity: The Special Theory}
(Amsterdam: North-Holland)

\endrefs

\end{document}